\documentclass[
showpacs,preprintnumbers,amsmath,amssymb]{revtex4}
\usepackage{amsmath}
\usepackage{empheq}
\usepackage{graphicx}
\begin{document}

\title{VISCOUS COUPLED FLUIDS IN INFLATIONARY COSMOLOGY}

\author{   I. Brevik$^{1}$\footnote{E-mail:iver.h.brevik@ntnu.no} and A. V. Timoshkin$^{2}${\footnote{E-mail:timoshkinAV@tspu.edu.ru}}}

\medskip

\affiliation{$^{1}$Department of Energy and Process Engineering, Norwegian University
of Science and Technology, N-7491 Trondheim, Norway}

\affiliation{$^{2}$Tomsk State Pedagogical University, Kievskaja Street 60, 634050 Tomsk, Russia;
Laboratory of Theoretical Cosmology, Tomsk State University of Control Systems and Radio-electronics, Lenin Avenue 40, 634050 Tomsk, Russia}

 \today

\begin{abstract}
We consider the inflation produced by two coupled fluids in the flat Friedmann-Robertson-Walker universe. Different cosmological models for describing  inflation by use of an inhomogeneous equation of state  for the fluid are investigated. The gravitational equations for  energy and  matter are solved, and  analytic representations for the Hubble parameter and the energy density are obtained.  Corrections in the energy density for  matter inducing the inflation and the coupling with energy are discussed. We  analyze the description of inflation induced by non-constant equation-of-state parameters from fluid viscosity. The correspondence between the spectral index and the tensor-to-scalar ratio recently  observed by the Planck satellite is considered.
\end{abstract}

\pacs{98.80.-k, 95.36.+x}
 \maketitle
\section{Introduction}

According to recent observational data the current  universe is accelerating \cite{1}. There exists an early-time accelerated epoch  after the Big Bang, called the "hot scenario" or inflationary period  (for reviews see Refs.~\cite{2,3}). During inflation  the total energy and the scale factor both increase exponentially  \cite{4}. This phenomenon suggests that the cosmic fluid has  properties quite different from standard matter and radiation. The simplest description of the accelerated expansion of the universe is given by the dark fluid  model (the Friedmann-Robertson-Walker (FRW) model), in which the cosmic fluid is taken to satisfy a homogeneous or alternatively an inhomogeneous   equation of state.

 Various problems associated with viscous cosmology have been  discussed  extensively in Refs.~\cite{weinberg71,padmanabham77,gron90,brevik94,brevik04,cataldo05,brevik06,brevik02,brevik10,brevik05,li09,brevik10a,sebastiani10,velten12,velten13,velten13a,bamba15}.  References \cite{nojiri05,capozziello06,nojiri07,nojiri06}  were devoted to inhomogeneous fluids, of which a time-dependent equation of state (EoS) in the presence of viscosity represents a sub-class. The behavior of a non-classical inflation was considered in Ref.~\cite{venkataratnam13}. Inhomogeneous fluid cosmology may also be understood as modified gravity (for  reviews, see Refs.~\cite{nojiri11,nojiri06a}), owing to the fact that it may be modeled  as a gravitational fluid with an inhomogeneous equation of state \cite{capozziello06a}. Then, it is to be expected  that following the unification of inflation with dark energy in modified gravity \cite{nojiri03}, one can achieve this unification in viscous cosmology also, including the presence of dark matter.

Some examples of inhomogeneous viscous coupled fluids were considered in Refs.~\cite{bamba12,elizalde14,brevik15}. In the papers \cite{myrzakulov14a,brevik14} a bounce cosmology induced by an inhomogeneous viscous fluid in FRW space-time was investigated. In Ref.~\cite{myrzakul13} different kinds of inhomogeneous viscous fluids were analyzed concerning the possibility to reproduce the current cosmic acceleration in flat  FRW space-time and the presence of finite-future time singularities, while in Ref.~\cite{brevik11} the noticeable  fact was proven that a viscous fluid with an inhomogeneous equation of state is  able to produce a Little Rip cosmology as a purely viscosity effect. Various cosmological models of the coupling between energy and matter were considered in the review \cite{bolotin15}.

In the present article we will investigate  cosmological models in the presence of  inflation. Some  variants of inhomogeneous viscous coupled fluids, of interest in the inflationary regime,  are considered. The influence of the interaction between energy and matter on the description of inflationary cosmology is analyzed. The slow roll parameters, the spectral index, and the tensor-to-scalar ratio  are calculated. Restrictions on the thermodynamic parameters, needed  to satisfy the Plank data, are obtained. The agreement between theoretical inflationary models and  observational data is discussed.

\section{	Viscous fluid models for inflation}

In this section we will consider the early-time universe, applying   the formalism of inhomogeneous viscous fluid in flat FRW  space-time. We will describe  the inflation in terms of the equation-of-state parameters, and the bulk viscosity. We will pay attention to the behavior of the Hubble parameter and the energy density in the beginning and in the end of the inflation.

We will start from a universe filled with two coupled fluids: energy and matter in a spatially flat  metric with scale factor $a$. The background equations are (see e.g., Ref.~\cite{nojiri05a}):
\begin{equation}
\left\{ \begin{array}{lll}
\dot{\rho}+3H(p+\rho)=-Q, \\
\dot{\rho}_1+3H(p_1+\rho_1)=Q, \\
\dot{H}=-\frac{k^2}{2}(p+\rho+p_1+\rho_1),
\end{array}
\right. \label{1}
\end{equation}
where $H=\dot{a}/a$ is the Hubble parameter and $k^2=8\pi G$,  $G$  being Newton's gravitational constant. Further, $p, \rho$ and $p_1,\rho_1$ are the pressure and the energy density of dark energy and dark matter correspondingly, and $Q$ is a function that accounts for the energy exchange between the fluids.  A dot denotes derivative with respect to cosmic time $t$. The cosmological constant $\Lambda$ is set equal to zero.

As is known, there  exists a number of studies on cosmology with multi-component viscous fluids, especially in order to explain the late-time cosmic acceleration - cf., for instance, Refs.~\cite{velten12,velten13,velten13a,brevik15,wang14,balakin11,balakin11a,nunes15}. It is then most natural  to allow for interaction terms between the components. This means physically that each component is to be regarded as a non-closed physical system, corresponding to an energy-momentum tensor that is not divergence-free in general. The requirement of  vanishing four-divergence can be made only on  the total energy-momentum tensor $T_{\rm Total}^{\mu\nu}$, i.e., $ {T_{\rm Total}^{\mu\nu}}_{;\nu}=0$. The new element in our analysis is to adopt a two-component model also in the inflationary region. One may ask: is there a physical reason for adopting this model in the very special inflationary epoch? To our knowledge, there is at present no basic reason. Rather, the model is introduced on phenomenological grounds, to make the theory also in the inflationary epoch similar to that commonly accepted in other epochs. It should here be emphasized that the state of acceleration is common for inflation and for the late universe, especially when the Big Rip is approached. In a future work we plan to extend the present analysis so as to unify inflation with dark energy, such that one component describes inflation where the other component effectively describes the dark energy epoch. In this way we hope to present a unified universe evolution with the use of a two-component fluid.

Let us now write down Friedmann's equation for the Hubble rate
\begin{equation}
H^2=\frac{k^2}{3}(\rho+\rho_1). \label{2}
\end{equation}
 We will assume that there is no relativistic matter $(\tilde{w}=0)$, and we will assume, as usual, that the dark matter  is dust.  Thus, $p_1=0$. We also put  $c=1$, as usual.

 The gravitational equation for matter reduces to
 \begin{equation}
 \dot{\rho}_1+3H\rho_1=Q. \label{3}
 \end{equation}
 We will now study some cases of inhomogeneous viscous coupled fluids in the  inflationary period.

\subsection{Fluid model with  $\omega =-\rho/(\rho+\rho_*)$,  and viscosity proportional to $H$ 	}

Let us  choose the viscosity to have the form \cite{myrzakulov14}
\begin{equation}
\zeta(H)=e^{-H/H_*}\, f(H), \label{4}
\end{equation}
where the star subscript refers to the end of the inflationary period. The function $f(H)$ will in the following be adjusted to have a physically reasonable form.		 Correspondingly, $H_*=H(t_*)$ refers to the end of inflation.
The function ({\ref 4}) is assumed to vary slowly with time, as inflation can be realized when the viscosity is slight.

Let us take the equation of state to have the following inhomogeneous form
\begin{equation}
p=\omega(\rho)\rho+\zeta(H), \label{5}
\end{equation}
where the thermodynamic parameter is  $\omega(\rho)=-\rho/(\rho+\rho_*)$ \cite{myrzakulov14}, and $\rho_*$ is the energy density at the end of inflation $(\rho_* \equiv H_*^2/k^2)$.

First of all, we will find the solution of the gravitational equation for the energy component \cite{capozziello06a}:
\begin{equation}
\dot{\rho}+3H\rho[1+\omega(\rho)]-3H^2\zeta(H)=-Q. \label{6}
\end{equation}
Let us suppose that the ratio $r=\rho_1/\rho$  is a constant, and rewrite Friedmann's equation (\ref{2}) in the form
\begin{equation}
\rho=\frac{3H^2}{k^2(1+r)}. \label{7}
\end{equation}
We will consider the following coupling between the fluids:
\begin{equation}
Q=9\delta H^3, \label{8}
\end{equation}
where $\delta$ is a positive dimensional constant. In geometric units, its dimension is $[\delta]={\rm cm}^{-2}$.

Since there exists no fundamental theory specifying the functional form of the coupling between the fluids, the present coupling model has necessarily to be phenomenological \cite{bolotin15}.

Further, we will analyze the model in the asymptotic limit corresponding to the initial stage of inflation,
\begin{equation}
\frac{H}{H_*} \ll 1. \label{9}
\end{equation}
Then $\zeta(H/H_*\ll1) = f(H)$, and the function $f(H)$ can be taken to be proportional to $H$,
\begin{equation}
f(H) =\theta H, \label{10}
\end{equation}
with $\theta=3/(k^2(1+r))$.

Taking into account Eqs.~(\ref{7},{8},{10}) the continuity equation (\ref{6}) for energy simplifies to
\begin{equation}
\frac{2}{3}\dot{H}+H^2\frac{\rho_*}{\theta H^2+\rho_*}=0. \label{11}
\end{equation}
As we are considering the initial stage of inflation, we may put the density $\rho_*$ in the form $\rho_*=\theta H_{\rm in}^2$ with $H_{\rm in}=H(t_{\rm in})$, where $t_{\rm in}$ is the initial time. In this approximation the solution of Eq.~(\ref{11}) becomes
\begin{equation}
H(t)=(\sqrt{\tau^2+1}-\tau)H_{\rm in}, \label{12}
\end{equation}
where $\tau=(3/4)H_{\rm in}(t-t_{\rm in})$. The Hubble parameter is positive, so we have an expanding universe.

Now, we solve the gravitational equation of motion (\ref{3}) for matter with the coupling terms (\ref{8}) and the Hubble parameter (\ref{12}),
\begin{equation}
\rho_1(t)=H^2\left[ 6\delta (e^\tau -1)+\frac{(\rho_1)_m}{H_{\rm in}^2}\right] \exp{(-2\tau  H/H_{\rm in})}. \label{13}
\end{equation}
Here $(\rho_m)_m=\rho_m(t_{in})$, and the time lies in the interval $t \in [t_{\rm in}, t_{\rm in}+4/(3H_{\rm in})]
$.
Finally, from Friedmann's  equation (\ref{2}) we find the energy density,
\begin{equation}
\rho(t)=H^2\left\{ \frac{3}{k^2}-\left[ 6\delta (e^\tau-1)+\frac{(\rho_1)_{\rm in}}{H_{\rm in}^2}\right] \exp{(-2\tau  H/H_{\rm in})}\right\}. \label{14}
\end{equation}
Thus we have shown how the inflation is realized, by using an inhomogeneous equation of state parameter and viscous coupled fluids.

Further, we can investigate how this inflationary model conforms with
 recent  results from the Planck satellite. We first calculate the  slow roll parameter $\varepsilon$,
\begin{equation}
\varepsilon=-\frac{\dot{H}}{H^2}=\frac{3}{4}\left( 1+\frac{\tau}{\sqrt{\tau^2+1}}\right). \label{15}
\end{equation}
Acceleration takes place at $\varepsilon <1$. This requirement is fulfilled, when  $t\in [t_{\rm in}, t_{\rm in}+\sqrt{2}/(3H_{\rm in})]$.  But the spectral parameters $n_s$ and $r$ become negative. Consequently, the results of the recent  BICEP2 experiments are not reproduced in the model.

\subsection{ Fluid model with $\omega(\rho)=\omega_0$   constant and viscosity proportional to $H^2$ }

In this section we will assume that the thermodynamic parameter  $\omega(\rho)=\omega_0$ is a constant in Eq.~(\ref{5}).
We  study the initial stage of inflation and take the function  $f(H)$ to be  proportional to the square of $H$,
\begin{equation}
f(H)=\tilde{\theta}H^2, \label{16}
\end{equation}
where  $\tilde{\theta} =\theta/3$.  As in the previous section the interacting term  $Q$  has the same form as in Eq.~(\ref{8}).

The gravitational equation of motion (\ref{6}) takes the form:
\begin{equation}
\frac{2}{3}\dot{H}+\left(\omega_0+\frac{\delta}{\tilde{\theta}}\right)H^2=0. \label{17}
\end{equation}
The Hubble parameter is equal to
\begin{equation}
H=\frac{2}{3\left( \omega_0+  \frac{\delta}{\tilde{\theta}}\right)(t-t_{\rm in})-\frac{2}{H_{\rm in}}}. \label{18}
\end{equation}
We  find the solution of the gravitational equation of motion for  matter as
\begin{equation}
\rho_1(t)=\left(\frac{H}{H_{\rm in}}\right)^{\frac{2}{\omega_0+
\delta/\tilde{\theta}}}
\left(\tilde{\rho}_{\rm in}+\frac{3\delta}{\omega_0-1+\frac{\delta
}{\tilde{\theta}}}H_{\rm in}^2\right)-
\frac{3\delta}{\omega_0-1+\frac{\delta}{\tilde{\theta}}}H^2, \label{19}
\end{equation}
where $\tilde{\rho}_{\rm in}=\rho_1(t_{\rm in})$ is the energy density of matter in the beginning of inflation.

The dark energy density is given by the expression
\begin{equation}
\rho(t)=3\left(\frac{1}{k^2}+\frac{\delta}{\omega_0+1+\frac{\delta}{\tilde{\theta}}}\right)H^2
 -\left(\frac{H}{H_{\rm in}}\right)^{\frac{2}{\omega_0+\delta/\tilde{\theta}}}
 \left( \tilde{\rho}_{\rm in}+\frac{3\delta}{\omega_0-1+\frac{\delta}{\tilde{\theta}}}H_{\rm in}^2\right). \label{20}
\end{equation}
Thus, we have constructed an example of  an  inhomogeneous viscous fluid cosmology coupled with matter,  applied to inflation.

Further, we will use the solution (\ref{18}) to calculate the slow roll parameter in the inflation,
\begin{equation}
\varepsilon=\frac{3}{2}\left( \omega_0+\frac{\delta}{\tilde{\theta}}\right), \label{21}
\end{equation}
where $\omega_0+\delta/\tilde{\theta}>0.$ As above, in order to have a regime of acceleration, one must require $\varepsilon <1$, thus $\omega_0<(2/3)-\delta/\tilde{\theta}$.

Another important slow roll parameter is \cite{myrzakulov14}
\begin{equation}
\eta=\varepsilon-\frac{1}{2\varepsilon H}\dot{\varepsilon}. \label{22}
\end{equation}

In our case, $\varepsilon =\eta$. The power spectrum is \cite{myrzakulov14}
\begin{equation}
\Delta_R^2=\frac{k^2H^2}{8\pi^2\varepsilon}. \label{23}
\end{equation}
Using Eq.~(\ref{18}) for the Hubble parameter, the power spectrum takes the form
\begin{equation}
\Delta_R^2=\frac{k^2}{3\pi^2\left( \omega_0+\frac{\delta}{\tilde{\theta}}\right) \left[3\left(\omega_0+\frac{\delta}{\tilde{\theta}}\right)(t-t_{\rm in})-\frac{2}{H_{\rm in}}\right]^2}. \label{24}
\end{equation}
From the slow roll parameters one can calculate the spectral index $n_s$  and the tensor-to-scalar ratio $r$, 	
\begin{equation}
n_s=1-6\varepsilon+2\eta, \quad r=16\varepsilon. \label{25}
\end{equation}
We obtain
\begin{equation}
n_s=1-6\left(\omega_0+\frac{\delta}{\tilde{\theta}}\right), \quad r=24\left(\omega_0+\frac{\delta}{\tilde{\theta}}\right). \label{26}
\end{equation}
In the particular case $\delta=\tilde{\theta}$, we obtain the same values for the parameters (\ref{21}), (\ref{22}), and (\ref{26}) as in Ref.~\cite{myrzakulov14}.

From the observations by the Planck satellite it is known that $n_s=0.9603\pm 0.0073$.
 In order to satisfy this result, we must require  $\omega_0+\delta/\tilde{\theta} \approx 0.00(6)$.  Consequently, the present model can produce  inflation.

\section{	Quasi-de Sitter expansion for inflation produced by two coupled fluids}

In this section  we will investigate a nonviscous model for the cosmic fluid. The equation of state is taken to have the following inhomogeneous form \cite{myrzakulov14},
\begin{equation}
\omega(\rho)=-1+a_1\rho^{1/2}-a_2\rho^{-1/2}, \label{27}
\end{equation}
where $a_1$ and $a_2$ are positive dimensional constants. This model describes a quasi-de Sitter inflationary expansion. We will find an exact solution of this model taking into account the coupling of two different fluids.

Let us start from  the gravitational equation for  energy,
\begin{equation}
 \dot{\rho}+3H\rho^{1/2}(a_1\rho-a_2)=-Q. \label{28}
 \end{equation}
We take the energy exchange between energy and matter in the form
 \begin{equation}
 Q=3\tilde{\delta}H^2, \label{29}
 \end{equation}
 where $\tilde{\delta}$ is a positive dimensional constant, and we put the constants $a_1$ and $a_2$ in Eq.~(\ref{27}) equal to
 \begin{equation}
 a_1=\frac{1}{\sqrt{3\theta}}, \quad a_2=\frac{\tilde{\delta}}{\sqrt{3\theta}}. \label{30}
 \end{equation}
 Taking into account Eqs.~(\ref{7}), (\ref{20}) and (\ref{29}) we obtain the continuation equation for energy (\ref{28}) in the simple form
 \begin{equation}
 2\dot{H}+\frac{3}{\sqrt{\theta}}H^3=0. \label{31}
 \end{equation}
 The solution gives the Hubble parameter in the form
 \begin{equation}
 H=\frac{1}{\sqrt{\frac{3}{\sqrt{\theta}}(t-t_{\rm in})+\frac{1}{H_{\rm in}^2}}}. \label{32}
 \end{equation}
Further, we find the solution for the gravitational equation of matter,
\begin{equation}
\rho_1(t)=3\tilde{\rho}\sqrt{\theta}
\, e^{-\frac{2\sqrt{\theta}}{H}}
\left[ C+2\,{\rm Ei}\left( \frac{2\sqrt{\theta}}{H}\right) \right],
\label{33}
\end{equation}
where ${\rm Ei}(2\sqrt{\theta}/H)$ is the integral exponential function and $C$ an arbitrary constant.
where   is the integral exponential function and   is the arbitrary constant.

Then the expression for the energy density becomes
\begin{equation}
\rho(t)= 3\left\{  \frac{H^2}{k^2}- \tilde{\rho}\sqrt{\theta}
\, e^{-\frac{2\sqrt{\theta}}{H}}
\left[ C+2\,{\rm Ei}\left( \frac{2\sqrt{\theta}}{H}\right) \right] \right\}. \label{34}
\end{equation}
Thus we have obtained a situation in which the equation-of-state parameter changes slowly during the inflationary period, taking into account the coupling of two fluids.

As in the previous section, we will consider how this inflationary model confirms with the Planck observational data. First, let us calculate the slow roll parameters $\varepsilon$ and $\eta$. From the solution (\ref{32}) we get
\begin{equation}
\varepsilon =\frac{3}{2\sqrt{\theta}}H, \quad \eta=\frac{3}{2}\varepsilon, \label{35}
\end{equation}
and the power spectrum is given by the expression
\begin{equation}
\Delta_R^2=\frac{k^2\sqrt{\theta}}{12\pi^2\sqrt{\frac{3}{\sqrt{\theta}}(t-t_{\rm in})+\frac{1}{H_{\rm in}^2}}}. \label{36}
\end{equation}
The spectral index $n_s$ and the tensor-to-scalar ratio $r$ are given by
\begin{equation}
n_s=1-3\varepsilon, \quad r=16\varepsilon. \label{37}
\end{equation}
For a de Sitter expansion we may introduce the number of  $N$-folds,
\begin{equation}
N=\int_{t_{\rm in}}^{t_*}H(t)dt. \label{38}
\end{equation}
In our case
\begin{equation}
N=\frac{2\sqrt{\theta}}{3}\left( \frac{1}{H_*}-\frac{1}{H_{\rm in}}\right). \label{39}
\end{equation}
The inflation is viable if $N>76$. Therefore, the slow roll parameters change during inflation,
\begin{equation}
\varepsilon = \frac{1}{N} \ll 1, \quad \eta = \frac{3}{2N} \ll 1, \label{40}
\end{equation}
and the slow roll conditions are satisfied.

The spectral indices are given by
\begin{equation}
n_s=1-\frac{3}{N}, \quad r=\frac{16}{N}. \label{41}
\end{equation}
For  $N=76$ the indices are $n_s=0.9605$ and $r=0.221$. Consequently, the results from the Planck satellite data can be realized from this model. According to the recent Planck data \cite{planck} the tensor-to-scalar ration is constrained to be $r<0.11$ (95\% CL).

\section{	Conclusion}

In the present paper we  investigated  coupled-fluid cosmological models  which take into account  viscosity properties of the fluid in  FRW flat space-time,  in a hot universe.  We  considered the influence of the interaction between  energy and matter during the inflationary very early stages of the evolution. We paid attention especially to  the initial stage of the inflation. Studying  the agreement between the theoretical inflationary models and the last results of the Planck satellite data, we showed that some restrictions on the thermodynamic parameters allowed satisfactory  correspondence with the observations.

Inhomogeneous fluid cosmology may be  understood as some kind of modified gravity (for a review, see Ref.~\cite{brevik11}), owing to the fact that it may be presented as a gravitational fluid with an inhomogeneous equation of state \cite{nojiri05a}. Then, it  follows that the unification of inflation with dark energy in modified gravity \cite{brevik11}  one can also achieve this unification in viscous cosmology, with the inclusion   of dark matter.

Our theory may be extended for the case of multiple coupled viscous fluids. The calculation may be done in the same way as above.

\bigskip

\section*{Acknowledgements}
              This work was supported by a grant from the Russian Ministry of Education and Science, project TSPU-139 (A.V.T.). We also thank Sergei D. Odintsov for valuable information.

\end{document}